\documentclass[12pt]{iopart}
\usepackage{iopams}
\usepackage{graphicx}
\begin{document}

\title[Subtleties of the clock retardation]{Subtleties of the clock retardation}
\author{D V Red\v zi\' c}

\address {Faculty of Physics, University of Belgrade, PO
Box 44, 11000 Beograd, Serbia} \eads{\mailto{redzic@ff.bg.ac.rs}}

\begin{abstract}
For a simple electromagnetic model of a clock introduced by
Jefimenko (clock $\#$ 1 in 1996 {\it Am. J. Phys.} {\bf 64} 812), a
change of the rate of the clock when it is set in uniform motion is
calculated exactly, employing the correct equation of motion of a
charged particle in the electromagnetic field and the universal
boostability assumption. Thus, for the clock under consideration, a
dynamical content of the clock retardation is demonstrated. Somewhat
surprisingly, the analysis presented discloses that some familiar
relativistic generalisations concerning the retardation of clocks
have to be amended.
\end{abstract}
%\maketitle

\section{Introduction}
Consider a thin massive ring of radius $a$ at rest in the laboratory
frame $S$, carrying a uniformly distributed charge $q_1$. Let the
axis of the ring be the $x$ axis, with the origin at the centre of
the ring. The electrostatic field of the ring on its axis is

\begin {equation}
\bi E(x,0,0) = \frac{\kappa q_1x \hat {\bi x}}{(a^2 + x^2)^{3/2}}\,
,
\end {equation}
where $\kappa \equiv 1/4\pi\epsilon_0$. A point charge $q_2$ of
opposite sign ($q_1q_2 < 0$), whose mass is $m$, is set at the point
on the positive $x$ axis with $x = \cal A$. If the charge $q_2$ is
released with zero initial velocity to move under the action of the
electrostatic field of the ring, it will oscillate along the axis
between the points $x = \pm\cal A$. The system `charged ring and
$q_2$' thus constitutes a primitive electromagnetic clock (clock
$\#$ 1 in \cite{OJ1}); denote the period of oscillations of $q_2$
for the clock at rest by $T_0$.

Assume now that the ring and the point charge are set in motion with
constant velocity $\bi v_0 = v_0\hat {\bi x}$ along the positive $x$
axis. What is the period of the clock in uniform motion?

The problem was posed by Jefimenko, who solved it by a direct
calculation of the period in the laboratory frame, employing
Heaviside's formulae for the electric and magnetic fields of a
uniformly moving point charge and the {\it longitudinal mass} of
$q_2$ \cite{OJ1}. The author found that the period of oscillations
of $q_2$ for the moving clock is

\begin {equation}
T_M = \frac{T_0}{\sqrt {1 - v_0^2/c^2}}\, .
\end {equation}
where $c^2 \equiv 1/\epsilon_0\mu_0$; $c$ is the speed of
electromagnetic waves {\it in vacuo} and at the same time the speed
of light {\it in vacuo}. Thus the clock `consisting of the charged
ring and the point charge runs {\it slower} when the clock is
moving, and the rate of the moving clock is $(1 - v_0^2/c^2)^{-1/2}$
times the rate of the same stationary clock', as predicted by
Special Relativity. Since the conclusion was reached via a dynamical
argument only, Jefimenko inferred that, for the clock under
consideration, his calculations `provide a dynamic
cause--and--effect type explanation of time dilation' \cite{OJ1}.
Moreover, pursuing this line of reasoning, he noted that `if the
slow rate of moving clocks can be indeed explained as a dynamic
cause--and--effect phenomenon rather than as the kinematic effect
enunciated by Einstein, then the slow rate of moving clocks cannot
be interpreted as a proof of time dilation' \cite{OJ2}.

Unfortunately, Jefimenko's argument involves a confusing assumption
that the velocity $v_0$ of the moving clock is much larger than the
maximum velocity of $q_2$ relative to the ring; moreover, the
desired conclusion (2) is reached due to the additional
approximation of {\it small} oscillations, i. e., $\cal A \ll$ $ a$.
Thus, while Jefimenko correctly pointed out a dynamical content of
the clock retardation, his approach masks some important aspects of
the phenomenon, making some of his inferences fallacious.

In this paper, an exact dynamical analysis of Jefimenko's clock $\#$
1 is given. It is demonstrated that if the clock in motion is to be
{\it relativistically valid}, i. e. to serve as an {\it identical}
standard of time also for a co-moving inertial observer, analysis of
its clockwork requires {\it two} frames, the lab frame $S$ and the
rest frame of the clock $S'$. Thus, contrary to Jefimenko's claim,
one cannot avoid Einstein's principle approach to Special
Relativity; for understanding of the clockwork, a blend of
constructive `dynamical' and principle approaches is needed.
Moreover, the analysis presented reveals that for this
`longitudinal' relativistic clock, while equation (2) still applies,
the clock retardation exhibits non-uniformity, which seems to be
overlooked in the literature.

Namely, Einstein in 1905 inferred from the Lorentz transformation,
for a (practically) {\it point clock}, that a clock travelling with
velocity $v$, `when viewed from the stationary system', runs slower
by the factor $(1 - v^2/c^2)^{-1/2}$ than the same clock at rest in
the stationary system (\cite{AE1}, cf also \cite{AE2}). Later, he
generalised this statement to a clock with a second hand and even
declared that `every happening in a physical system slows down when
the system is set into translational motion' \cite{AE3}. (M{\o}ller
paraphrased this by stating that `any physical system which is
moving relative to a system of inertia must have a slower course of
development than the same system at rest' \cite{CM1}.) Thus,
according to Einstein, not only moving clocks run slow but time
itself is `dilated' in moving systems\footnote[1] {The somewhat
misleading term `time dilation' was probably introduced by Tolman
\cite{RT}; incidentally, it appears that `dilatation' would be the
{\it gramatically} correct variant \cite{FWS}.}, `but this slowing
occurs only from the standpoint of a non--comoving coordinate system
(observer)' \cite{AE3}. However, our analysis of Jefimenko's
`longitudinal' clock demonstrates that Einstein's familiar
generalisation about slowing down of processes in a moving physical
system need to be amended. Generally, some happenings in a physical
system speed up when the system is set into translational motion.
Also, for a periodic process, in a physical system travelling with a
velocity $v_0$ relative to a system of inertia, while equation (2)
applies, generally it is not true that development of the process
has a slower course by the factor $(1 - v_0^2/c^2)^{-1/2}$ than in
the same system at rest: the clock retardation may be non-uniform.
In this paper, a dynamical content of the phenomenon is illustrated
by a detailed examination of Jefimenko's clock $\#$ 1
\cite{OJ1,OJ2}.

\section{Direct calculation of the retardation of Jefimenko's clock }

\noindent Now we shall calculate exactly the period of Jefimenko's
clock described in Introduction, when it is at rest in the lab frame
$S$, and when it is in uniform motion with velocity $\bi v_0 =
v_0\hat {\bi x}$ along the $x$ axis (the axis of the ring) with
respect to the lab. As is well known, in order to be {\it
relativistically valid}, a clock must operate according to some
Lorentz--covariant laws. This can be fulfilled for Jefimenko's
clock, since its operation is based on Maxwell's equations (which
can be made to be Lorentz--covariant (cf, e.g., \cite{WGVR},
\cite{DVR1})) and the equation of motion of a charge $q^*$ in the
electromagnetic field

\begin {equation}
\frac{\rmd}{\rmd t}\left(\frac {m^*\bi v}{\sqrt {1-v^2/c^2}}\right)
= q^* \bi E + q^* \bi v \times \bi B\, ,
\end {equation}
where $m^*$ is the mass of the charge, $\bi v$ is its instantaneous
velocity, $\bi E$ is the electric field and $\bi B$ is the magnetic
flux density; the time parameter $t$ in the $S$ frame is interpreted
in the standard way, employing propagation of light signals {\it in
vacuo} as the absolute {\it time keeper}, assuming that Einstein's
clock synchronisation is a valid procedure \cite{CM1}. Equation (3)
fits the experimental facts if the {\it additional independent}
assumption that $m^*$ is time-independent is introduced; with that
assumption, equation (3) can be Lorentz--covariant too \cite{DVR2}.

\subsection{Clock at rest}
\noindent In this subsection we discuss the clockwork of Jefimenko's
clock at rest.

The correct equation of motion of the charge $q_2$ in an electric
field is

\begin {equation}
m\frac{\rmd}{\rmd t}\left(\frac {\bi v}{\sqrt {1-v^2/c^2}}\right) =
q_2 \bi E \, ,
\end {equation}
where the mass $m$ of $q_2$ is assumed to be time--independent.
Equation (4) and identity

\begin {equation}
\bi v \cdot\frac{\rmd}{\rmd t}\left(\frac {\bi v}{\sqrt
{1-v^2/c^2}}\right) \equiv c^2\frac{d}{dt}\left(\frac {1}{\sqrt
{1-v^2/c^2}}\right)\, ,
\end {equation}
imply that

\begin {equation}
mc^2\frac{\rmd}{\rmd t}\left(\frac {1}{\sqrt {1-v^2/c^2}}\right) =
q_2\bi E\cdot \bi v\, .
\end {equation}
Specifying to our problem, $\bi E$ is the electrostatic field of the
ring on its axis given by equation (1), and $\bi v = v_x\hat {\bi
x}$, since the motion is along the $x$ axis. Using equations (4),
(6) and (1) one obtains\footnote[2] {The force $q_2\bi E$ is always
parallel to the instantaneous velocity $\bi v$ of the charge $q_2$
so that equation (7) can be derived using the concept of
`longitudinal' mass, taking into account that the Lorentz force
expression is a {\it pure} force (cf \cite{WR,DVR2}); I preferred
not to employ here the potentially misleading concept of
`longitudinal' mass.}

\begin {equation}
\frac {\rmd v_x}{\rmd t} = -\frac {\kappa |q_1q_2|}{m}\left ( 1 -
\frac {v_x^2}{c^2}\right )^{3/2}\frac {x }{(a^2 + x^2)^{3/2}}\, ,
\end {equation}
which can obviously recast into

\begin {equation}
\frac {\rmd v_x}{\rmd x}v_x = -\frac {\kappa |q_1q_2|}{m}\left ( 1 -
\frac {v_x^2}{c^2}\right )^{3/2}\frac {x }{(a^2 + x^2)^{3/2}}\, .
\end {equation}
Separating variables and integrating, setting $v_x = 0$ when $x =
\cal A$ and solving for $v_x$ yields
\begin{equation}
v_x = \frac{\rmd x}{\rmd t} = \mp c \{...\}^{1/2}\, ,
\end{equation}
where $-$ and $+$ sign corresponds to the motion of $q_2$ in the
direction of decreasing $x$ and increasing $x$, respectively, and
\begin{equation}
\fl \{...\} \equiv \{1 - [1+ (\kappa|q_1q_2|/mc^2)(1/\sqrt {a^2 +
x^2} - 1/\sqrt {a^2+ {\cal A}^2})]^{-2} \} \, .
\end{equation}
Equation (9) implies that, for the oscillator at rest, passage of
$q_2$ from $x$ to $x + \rmd x$ lasts time interval
\begin{equation}
\rmd t = \mp (1/c)\{...\}^{-1/2}\rmd x\, .
\end{equation}
Thus, the period $T_0$ of the oscillator at rest is given
by\footnote[3] {Incidentally, in the case of {\it small}
oscillations, i.e. when ${\cal A} \ll a$, from equation (12) one
obtains that

$$
T_0 \approx \frac{2}{c} \int_{-\cal A}^{\cal A} \{1 - [1+ ({\cal
K}/2mc^2)({\cal A}^2 - x^2)]^{-2} \}^{-1/2}\rmd x\, .
$$
where ${\cal K} \equiv \kappa |q_1q_2| /a^3$. Note that when ${\cal
K} {\cal A}^2 \ll mc^2$, from the last equation one obtains the
familiar expression for the period of the simple harmonic
oscillator, $T_0 = 2\pi \sqrt {m/{\cal K}}$ (cf also \cite{CM2},
\cite{DVR3}).}

\begin{equation}
T_0 = \frac{2}{c} \int_{-\cal A}^{\cal A} \{... \}^{-1/2}\rmd x\, .
\end{equation}

\subsection{Clock in uniform motion}
\noindent Assume now that the same clock is set in uniform motion
along the $x$ axis (the axis of the ring) with constant velocity
$\bi v_0 = v_0\hat {\bi x}$, so as to be relativistically valid, i.
e. to serve as an identical standard of time also for a co-moving
observer. The $x$ coordinate of the charge $q_2$, $x$, can be
expressed as

\begin{equation}
x = x_c + x_r\, ,
\end{equation}
where $x_c$ is the $x$ coordinate of the centre of the ring, and
$x_r \equiv x - x_c$ is the relative coordinate of $q_2$ with
respect to the instantaneous position of the centre. Since $v_x =
\rmd x/\rmd t$ and $v_0 = \rmd x_c/\rmd t$, one has

\begin{equation}
v_x = v_0 + v_{xr}\, ,
\end{equation}
where $v_{xr} = \rmd x_r/\rmd t$ is the relative velocity of $q_2$
with respect to the centre of the ring, {\it measured in the lab
frame} $S$.

The charge $q_2$ moves under the action of the electromagnetic field
on the axis of the uniformly moving charged ring (the ring
constitutes the framework of Jefimenko's clock). The electromagnetic
field is calculated using the formulae for the $\bi E$ and $\bi B$
fields of a point charge $q$ moving with constant velocity $\bi
v_0$, that were first obtained by Heaviside \cite{OH1,OH2} (the $\bi
B$ field was rederived by J J Thomson \cite {JJT}, cf Jefimenko
\cite{OJ3} and references therein), long before the advent of
Special Relativity. The electric field of $q$ (radial but not
spherically symmetrical) is given by

\begin {equation}
\bi E(\bi r,t) = \frac {\kappa q \bi r}{r^3}\frac{1 - v_0^2/c^2}{(1
- v_0^2\sin^2\theta/c^2)^{3/2}}\, ,
\end {equation}
where $\bi r$ is the position vector of a field point with respect
to the instantaneous (at the {\it same} instant t) position of $q$,
$\theta$ is the angle between $\bi r$ and the velocity $\bi v_0$,
and $c^2 \equiv 1/\epsilon_0\mu_0$. (Recall that throughout the
relativity paper \cite{AE1}, Einstein used the same symbol ($V$) for
the speed of light {\it in vacuo} and the speed of electromagnetic
waves {\it in vacuo} ($V \equiv 1/\sqrt{\epsilon_0\mu_0}$ in SI
system, not used by Einstein, he employed the Heaviside--Lorentz
units), linking thus Special Relativity with Maxwell's theory (cf
\cite{DVR4}, p 197).) The magnetic flux density is

\begin {equation}
\bi B(\bi r,t) = \epsilon_0\mu_0 \bi v_0 \times \bi E(\bi r,t)\, .
\end {equation}

A simple analysis employing formula (15) gives the following
expression for the electric field on the axis of the moving charged
ring

\begin {equation}
\bi E_M = \frac{\kappa q_1 (1 - v_0^2/c^2) x_r\hat {\bi x}}{(a^2 +
x_r^2 - a^2v_0^2/c^2)^{3/2}}\, ,
\end {equation}
as Jefimenko pointed out; the subscript $M$ serves as a reminder
that the field is due to the moving ring--charge. The $\bi B$ field
on the axis of the ring obviously vanishes. Using equations (4), (6)
and (17) one obtains

\begin {equation}
\frac {\rmd v_x}{\rmd t} = -\frac {\kappa |q_1q_2|}{m}\left ( 1 -
\frac {v_x^2}{c^2}\right )^{3/2}\frac {(1 - v_0^2/c^2) x_r}{(a^2 +
x_r^2 - a^2v_0^2/c^2)^{3/2}}\, .
\end {equation}
Using equation (14), the last equation can obviously be recast into

\begin {equation}
\frac {\rmd v_x(v_x - v_0)}{(1 - v_x^2/c^2)^{3/2}} = -\frac {\kappa
|q_1q_2|}{m}\frac {x_r\rmd x_r(1 - v_0^2/c^2)}{[a^2(1 - v_0^2/c^2) +
x_r^2]^{3/2}}\, .
\end {equation}

Now we have to specify our clock so as to be {\it relativistically
valid}. As can be seen, this cannot be done without recourse to
Einstein's principle approach to Special Relativity, i. e. without
employing the Lorentz transformation and the universal boostability
assumption (cf \cite{DVR3,DVR4}). A little reflexion reveals that
one has to choose initial condition $v_x = v_0$ for $x_r = {\cal A}
\sqrt {1 - v_0^2/c^2}$. A glance at equation (18) shows that the
charge $q_2$ will oscillate between the points $x_r = \pm {\cal A}
\sqrt {1 - v_0^2/c^2}$, where $v_x = v_0$.

Integration of equation (19), taking into account the initial
condition, gives

\begin {equation}
\frac {1 - v_0v_x/c^2}{\sqrt {1 - v_x^2/c^2}} \frac {1 }{\sqrt {1 -
v_0^2/c^2}} = 1 + \gamma\, ,
\end {equation}
where

\begin {equation}
\gamma \equiv  \frac {\kappa |q_1q_2|}{mc^2}\left (\frac {1 }{\sqrt
{a^2 + x_r^2/(1 - v_0^2/c^2)}} - \frac {1 }{\sqrt {a^2 + {\cal A}^2
}}\right )\, .
\end {equation}
Solving for $v_x$ and using equation (14), after a somewhat lengthy
but in every step simple calculation, skipping details, one obtains
for $v_{xr}$

\begin {equation}
v_{xr} = \frac {\rmd x_r}{\rmd t} = \mp \frac {c(1 - v_0^2/c^2)\sqrt
{1 - 1/(1 + \gamma)^2}}{1 \mp (v_0/c)\sqrt {1 - 1/(1 + \gamma)^2}}\,
,
\end {equation}
where $-$ and $+$ sign corresponds to the motion of $q_2$ in the
direction of decreasing $x_r$ and increasing $x_r$, respectively.
Equation (22) obviously implies that passage of $q_2$ from $x_r$ to
$x_r + \rmd x_r$ lasts time interval

\begin {equation}
\rmd t =  \mp \frac {1 \mp (v_0/c)\sqrt {1 - 1/(1 + \gamma)^2}}{c(1
- v_0^2/c^2)\sqrt {1 - 1/(1 + \gamma)^2}}\rmd x_r\, ,
\end {equation}
Introducing

\begin {equation}
x_r^* = \frac {x_r}{\sqrt {1-v_0^2/c^2}} \, ,
\end {equation}
and

\begin {equation}
\gamma^* = \gamma = \frac {\kappa |q_1q_2|}{mc^2}\left (\frac {1
}{\sqrt {a^2 + {x_r^*}^2}} - \frac {1 }{\sqrt {a^2 + {\cal A}^2
}}\right )\, .
\end {equation}
equation (23) can be recast into

\begin {equation}
\rmd t =  \mp \frac {[1/\sqrt {1 - 1/(1 + \gamma^*)^2} \mp
v_0/c]}{\sqrt {1 - v_0^2/c^2}\,\,c}\rmd x_r^*\, ,
\end {equation}
or, equivalently,

\begin {equation}
\rmd t =  \mp \frac {1}{\sqrt {1 - v_0^2/c^2}} \,\frac {1}{c}\,
[\{...^*\}^{-1/2} \mp v_0/c]\rmd x_r^*\, ,
\end {equation}
where

\begin{equation}
\fl \{...^*\} \equiv \{1 - [1+ (\kappa|q_1q_2|/mc^2)(1/\sqrt {a^2 +
{x_r^*}^2} - 1/\sqrt {a^2+ {\cal A}^2})]^{-2} \} \, .
\end{equation}
Comparing equations (27) and (11), taking into account that $x_r^*$
runs from $\cal A$ to $- \cal A$ and {\it vice versa}, it follows
that clock retardation is non-uniform in the case of Jefimenko's
clock; simple slowing down by the factor $1/\sqrt {1 - v_0^2/c^2}$
is clearly violated in the `life' of this `longitudinal'
clock.\footnote[4] {This is in contradistinction to the case of a
`transverse' clock where the clock retardation is uniform. A
dynamical content of the clock retardation in the case of a simple
electromagnetic model of a `transverse' {\it relativistically valid}
clock was recently analysed exactly in \cite{DVR3}.} Particularly,
using equation (27) one finds that travelling of $q_2$ downwards
from $x_r = {\cal A } \sqrt {1 - v_0^2/c^2}$ to $x_r = - {\cal A }
\sqrt {1 - v_0^2/c^2}$ lasts time interval

\begin {equation}
(\Delta t)_{\rm {down}} = - \frac {1}{\sqrt {1 - v_0^2/c^2}} \,\frac
{1}{c}\, \left [\int_{\cal A}^{-\cal A}\{...^*\}^{-1/2}\rmd x_r^* -
\int_{\cal A}^{-\cal A}\frac {v_0} {c} \rmd x_r^*\right ]\, ,
\end {equation}
whereas the reverse travel upwards lasts

\begin {equation}
(\Delta t)_{\rm {up}} = \frac {1}{\sqrt {1 - v_0^2/c^2}} \,\frac
{1}{c}\, \left [\int_{-\cal A}^{\cal A}\{...^*\}^{-1/2}\rmd x_r^* +
\int_{-\cal A}^{\cal A}\frac {v_0} {c} \rmd x_r^*\right ]\, .
\end {equation}
Taking into account equation (12), these expressions can be recast
into

\begin {equation}
(\Delta t)_{\rm {down}} =  \frac {1}{\sqrt {1 - v_0^2/c^2}}  \left
(\frac {T_0} {2} - \frac {v_0} {c^2} 2 \cal A \right )\, ,
\end {equation}

\begin {equation}
(\Delta t)_{\rm {up}} =  \frac {1}{\sqrt {1 - v_0^2/c^2}}  \left
(\frac {T_0} {2} + \frac {v_0} {c^2} 2 \cal A \right )\, .
\end {equation}
Consequently, for the period of Jefimenko's uniformly moving
`longitudinal' clock, one obtains

\begin {equation}
T_M = (\Delta t)_{\rm {down}} + (\Delta t)_{\rm {up}} =  \frac
{T_0}{\sqrt {1 - v_0^2/c^2}}\, ,
\end {equation}
as expected for a clock the clockwork of which is based on some
Lorentz--covariant (or that can be made to be Lorentz--covariant,
cf, e. g., \cite{WGVR,DVR1}) laws; the period of the clock in motion
is by the factor $1/\sqrt {1 - v_0^2/c^2}$ greater than the period
of the {\it same} clock at rest, all with respect to the lab frame
$S$. However, the present analysis reveals a dynamical content of
the phenomenon and its non-uniform character, which are hidden in
the conventional `kinematical' approach.

\section {Discussion}
The above analysis of Jefimenko's `longitudinal' clock demonstrates
that the clock retardation is basically of a dynamical origin, a
consequence of velocity--dependence of the forces governing the
operation of the clock. However, it also demonstrates that an exact
{\it one frame} derivation of the clock retardation for this
`longitudinal' type of clock is not possible; in order to specify
the clock so as to be {\it relativistically valid} ({\it inter
alia}, to choose adequately the amplitude of oscillations of $q_2$
for the clock in motion), one has to recourse to Einstein's
principle approach to Special Relativity.\footnote[5] {This is
analogous to the case of a `transverse' clock, where it seems that
an exact {\it one frame} derivation of the clock retardation is
impossible too (cf \cite {DVR3}).} Thus, perhaps somewhat
surprisingly for adherents of a constructive dynamical approach to
the theory, it appears that only a blend of constructive dynamical
and principle approaches leads to a complete insight into phenomena.

Another outcome of our analysis of Jefimenko's clock is that some
familiar relativistic generalisations concerning time need to be
amended. Notably, declarations such as `any physical system which is
moving relative to a system of inertia must have a slower course of
development than the same system at rest' \cite{CM1}, prove to be
fallacious in the general case.\footnote[6] {M\"{o}ller's statement
is clearly wrong in the case of the `longitudinal' light--pulse
clock (cf, e. g., \cite{APF}, pp 105--9). Namely, a simple analysis
reveals that the `downwards' segment of life of the moving light
clock (i. e., duration of the motion of light--pulse in the
direction opposite to the direction of motion of the clock itself)
lasts {\it shorter} than the corresponding segment of life of the
{\it same} clock at rest, all with respect to the lab frame $S$,
since

$$
\frac {l_0\sqrt{1 - v_0^2/c^2}} {c + v_0} < \frac {l_0} {c}\, ,
$$
where $l_0$ is the rest length of the clock and $v_0$ is its
velocity relative to $S$.} However, M\"{o}ller's statement is valid
in the special case of a {\it periodic complete} process occurring
in Jefimenko's clock. Indeed `moving clocks run slow' from the
standpoint of a non--comoving observer, but with addendum that
within one period this slowing is generally non-uniform.

The power and precision of Einstein's principle approach to Special
Relativity manifests itself in the universality of its results,
regardless of the type of Lorentz--covariant mechanism responsible
for a particular phenomenon. Therefore it is to be expected that
some of our basic conclusions concerning the properties of {\it
relativistically valid} clocks in uniform motion are reachable also
via the principle approach. This is indeed so, as the following
argument reveals.

Consider the standard Lorentz transformation

\begin {equation}
t =  \frac {t' + v_0x'/c^2}{\sqrt {1 - v_0^2/c^2}}\, , \quad x =
\frac {x' + v_0t'}{\sqrt {1 - v_0^2/c^2}}\, , \quad y =y'\, , \quad
z = z'\, ,
\end {equation}
where unprimed coordinates refer to the lab frame $S$ and primed
coordinates refer to an inertial frame $S'$ which is in standard
configuration with $S$ ($S'$ is uniformly moving at speed $v_0$
along the common positive $x,x'$-axes, and the $y$- and $z$-axis of
$S$ are parallel to the $y'$- and $z'$-axis of $S'$).

Let $S'$ be the rest frame of the {\it framework} of a {\it
relativistically valid} clock. For Einstein's (practically) point
clock, the first equation (34) and the condition $x' = const$ imply
the familiar result

\begin {equation}
\rmd t =  \frac {\rmd t'}{\sqrt {1 - v_0^2/c^2}}\, .
\end {equation}

Consider now a `transverse' relativistically valid clock, i. e. the
one the clockwork of which involves oscillations of a material point
with $x' = const$, whereas $y'$ and $z'$ are variable (cf
\cite{DVR3}). Clearly, equation (35) applies to this type of clock
too. Taking into account the principle of (special) relativity, and
the universal boostability assumption \cite{DVR4,DVR5}, equation
(35) can be given the following interpretation: {\it any} segment of
`life' of a `transverse' clock that is moving uniformly with the
velocity $v_0$ lasts longer by the factor  $1/\sqrt {1 - v_0^2/c^2}$
than the {\it same} segment of `life' of the {\it same} clock at
rest in $S$, all with respect to the lab frame $S$ \cite {DVR3}.

Finally, consider a `longitudinal' relativistically valid clock,
which involves oscillations of a material point with $x' \neq
const$, and $y'$ and $z'$ need not be constant. In this case, the
first equation (34) implies

\begin {equation}
\rmd t =  \frac {\rmd t' + (v_0/c^2)\rmd x'}{\sqrt {1 - v_0^2/c^2}}
= \frac {\rmd t' + (v_0/c^2)v_x'\rmd t'}{\sqrt {1 - v_0^2/c^2}}\, .
\end {equation}
where $v_x'$ is the $x'$ component of the instantaneous velocity of
the material point in $S'$. As can be seen, the first equation (36)
is tantamount to equation (27), taking into account the universal
boostability assumption. Obviously, equation (35) does not apply to
this `longitudinal' type of clock; the clock retardation now is
non-uniform, contrary to a general consensus in the literature that
equation (35) is universally valid. However, {\it any} clock
involves by definition a periodic process; thus, integration of
equation (36) over a period in $S'$ yields

\begin {equation}
T = \frac {T'}{\sqrt {1 - v_0^2/c^2}}\, ,
\end {equation}
where $T$ and $T'$ are the periods of the clock as observed in $S$
and $S'$, respectively, since $x'$ takes its initial value after the
period $T'$. As can be seen, equation (37) can be given the same
meaning as equation (33).\footnote[7] {But, we derived equation (33)
for Jefimenko's clock, which is strictly 'longitudinal' ($y' = z' =
0$), whereas the `kinematical' approach yields equation (37) for
more general 'longitudinal' clocks. Incidentally, any real clock
involves damping; thus, Jefimenko's clock represents an ideal clock
since the radiation reaction force was neglected in its clockwork.}

On the other hand, in the case of an {\it aperiodic} process
involving a point particle, no inference can be deduced from
equation (36), except that it is {\it not} generally true that
`every happening in a physical system slows down when the system is
set into translational motion [in a rest properties--preserving
way]' \cite{AE3}. This implies that particular instances of `time
retardation' must be carefully considered.

Note that the universality and elegance of the above `kinematical'
deductions concerning clocks in motion may be deceptive. Namely, one
should keep in mind that any clock retardation involves a complex
dynamical process, which is completely masked in the conventional
`kinematical' derivation.

One last point. {\it Basic} inferences concerning slowing down of
moving clocks from the standpoint of a non--comoving observer are
deduced from Einstein's two principles of Special Relativity, aided
with the universal boostability assumption. The present discussion
reveals that for subtler insights, dynamical considerations are
needed, as expected. Moreover, the dynamical approach indicates
where to look within the principle approach. On the other hand, the
principle approach anticipates simplicity and universality behind an
intricate dynamical mechanism; for instance, one knows in advance
that $T_M$ depends on dynamical parameters only through dependence
of $T_0$ on those parameters. Thus one can evade cumbersome
dynamical calculation for a physical system in uniform motion with
respect to the lab frame by combining much simpler dynamical
calculation in the rest frame of the system {\it and} the
transformation laws of relevant quantities known from the principle
approach, as Einstein noted long ago for `all problems in the optics
of moving bodies' \cite{AE1}.

In his excellent book {\it Special Relativity}, French warned to
beware of `glib statements involving relativity theory' \cite{APF}.
This paper, hopefully, represents such a warning, illustrating ever
present danger of unjustified generalisations.

\section*{Acknowledgments}
My work is supported by the Ministry of Science and Education of the
Republic of Serbia, project No. 171028.

\end{document}